\documentclass[prd,aps,preprint,superscriptaddress,showkeys,showpacs,
 preprintnumbers,amsmath,amssymb,floatfix]{revtex4-1}
\usepackage{graphicx}
\usepackage{epsfig}
\usepackage{amsfonts}
\usepackage{dsfont}
\usepackage{amsmath}
\usepackage{amssymb}
\usepackage[sort&compress]{natbib}
\usepackage{dcolumn}
\usepackage{multirow}
\usepackage{bigstrut}
\usepackage{mathrsfs}
\usepackage{type1cm}
\usepackage{placeins}
\usepackage{color}
\usepackage{rotating}
\usepackage{slashed}
\usepackage{subfigure}
\usepackage[mathscr]{euscript}
\DeclareMathOperator*{\SumInt}{%
\mathchoice%
  {\ooalign{$\displaystyle\sum$\cr\hidewidth$\displaystyle\int$\hidewidth\cr}}
  {\ooalign{\raisebox{.14\height}{\scalebox{.7}{$\textstyle\sum$}}\cr\hidewidth$\textstyle\int$\hidewidth\cr}}
  {\ooalign{\raisebox{.2\height}{\scalebox{.6}{$\scriptstyle\sum$}}\cr$\scriptstyle\int$\cr}}
  {\ooalign{\raisebox{.2\height}{\scalebox{.6}{$\scriptstyle\sum$}}\cr$\scriptstyle\int$\cr}}
}

\begin{document}

\preprint{\today}

\title{Cosmic-ray fermion decay by emission of on-shell $W$ bosons with CPT violation}

\author{D. Colladay}
\affiliation{New College of Florida, Sarasota, FL, 34243}

\author{J. P. Noordmans}
\author{R. Potting}
\affiliation{CENTRA, Departamento de F\'isica, Universidade do Algarve, 8005-139 Faro, Portugal}

\begin{abstract}
We study CPT and Lorentz violation in the electroweak gauge sector of the Standard Model in the context of the Standard-Model Extension.
In particular, we consider the Lorentz-violating and CPT-odd Chern-Simons like parameter for the $W$ boson, which is thus far unbounded by experiment.
We demonstrate that any non-zero value of this parameter implies that,
for sufficiently large energies,
one of the polarization modes of the $W$ boson propagates with spacelike four-momentum.
In this scenario, emission of $W$ bosons by ultra-high-energy cosmic rays is possible.
We calculate the induced fermion energy-loss rate and we deduce the first limit on the pertinent Lorentz- and CPT-violating parameter that couples to the $W$ boson.
Consistency between the quantum description in various reference frames is preserved by using a recently formulated covariant quantization procedure for massive photons and
applying it to the $W$ bosons.
\end{abstract}

\maketitle

Searches for departures from relativity are motivated by candidate theories of quantum gravity that allow for (spontaneous) Lorentz violation (LV) \cite{qgmodels}. Studies of LV, both theoretical and experimental, are facilitated by a general effective-field-theory framework called the Standard-Model Extension (SME) \cite{sme}. The Lagrangian of the matter sector of this framework contains all LV gauge-invariant effective operators that can be built from the conventional Standard-Model fields, coupled to vector and tensor coefficients that parametrize the LV. In fact, the SME also contains all CPT-violating operators, since in any local interacting quantum field theory CPT violation (CPTV) implies LV \cite{Gre02}. The SME thus enables a general quantification of the exactness of Lorentz and CPT symmetry in the form of observational contraints on the Lorentz-violating coefficients (LVCs)~\cite{datatables}. Ultimately, such restrictions on LV and CPTV can provide guidelines to find the correct theory of quantum gravity.

A possible observational consequence of LV, that can be addressed using astrophysical data, is vacuum Cherenkov radiation \cite{vacuumchernk}. The LVCs can in some cases be interpreted as inducing a refractive index for the vacuum. Consequently, the velocity of charged particles above some energy threshold might exceed the phase velocity of light. This causes these particles to rapidly lose energy through photon emission. The mere observation of high-energy cosmic particles can then be used to constrain the LVCs.

In this work, we consider a similar process, but with the emitted photon replaced by a $W$ boson.
We assume the latter to obey a LV and CPTV dispersion relation, originating from the superficially renormalizable part of the SME, called the minimal SME (mSME).
In this case, the LV originates from a Chern-Simons like addition to the Standard-Model Lagrangian \cite{Carroll} and is captured by one four-vector: $k^\mu_{2}$.
Such a theory has been shown to be consistently and covariantly quantizable, 
despite the presence of spacelike momenta,
which are necessary for vacuum Cherenkov radiation to occur. Although such momenta also give rise to negative-energy states in some (highly boosted) observer frames, the theory turns out to be stable within the framework of conventional quantum field theory \cite{covquant}.

Apart from the fact that the $W$-boson parameter $k^\mu_{2}$ has not been studied before, an essential difference between the analysis performed here 
and several previous calculations
in the literature involving massive photons \cite{vacuumchernk}, is that
the $W$-boson mass is very large compared to the incoming fermion mass.  
In previous studies, the mass of the photon was an arbitrarily small parameter (well below experimental bounds for the photon mass) that was only large enough
to dominate the ill effects of $k^\mu$. It was primarily introduced as a regulator to define the quantization
procedure and allow for consistent calculations. Here, the $W$-boson mass is large and therefore determines the size of the threshold energy (together with $k^\mu_{2}$).

The relevant LV and CPT-odd contribution to the Lagrangian is given by
\begin{equation}
\mathcal{L}_{\rm LV} = \frac{1}{2}(k_{2})_\kappa\epsilon^{\kappa\lambda\mu\nu}W^+_\lambda W^-_{\mu\nu} + {\rm h.c.} \ ,
\label{LVlagrangian}
\end{equation}
where $W^\pm_{\mu\nu} = \partial_\mu W^\pm_\nu - \partial_\nu W^\pm_\mu$ and $W^\pm_\mu$ represents the physical charged $W$ boson. The LV four-vector $k^\mu_{2}$ is real and can either be timelike, lightlike, or spacelike. It corresponds to the $SU(2)$ gauge-boson parameters defined in Ref.~\cite{sme}, which is complemented by a $U(1)$ gauge-boson parameter $k_1^\mu$.
The four-vectors $k_1^\mu$ and $k_2^\mu$ parametrize all the possible CPTV in the $SU(2)\times U(1)$ gauge sector of the mSME \cite{kphicontr},
with the exception of a $k_0^\mu$ parameter coupled to a term linear in the $U(1)$ gauge field. 
This term is mostly ignored, since it generates a linear instability in the potential.
In addition to the term in Eq.~\eqref{LVlagrangian},
other terms exist that contain combinations of $k_1^\mu$ and $k_2^\mu$. 
These include a $Z$-boson term analogous to Eq.~\eqref{LVlagrangian},
two gauge-boson three-point vertices, and a mixing term involving the $Z$-boson and the photon \cite{mixingnote}.
We will ignore all of these in the following.
Furthermore, there is a well-studied photon term, analogous to Eq.~\eqref{LVlagrangian},
coupled to a LVC $k_{AF}^\mu = 2\cos^2\theta_w k_1^\mu + \sin^2\theta_w k_2^\mu$, with $\theta_w$ the weak mixing angle,
which is experimentally limited to be smaller than $10^{-43}\,\mathrm{GeV}$ \cite{datatables}.
For our present purposes,
we can thus consider this linear combination of $k_1^\mu$ and $k_2^\mu$ to be zero.

We aim to calculate the rate at which a Dirac fermion with mass $m_1$ decays
to a LV $W$ boson with mass $M$ and a Dirac fermion with mass $m_2$,
using results from \cite{covquant} that will allow for a consistent
interpretation of the rate in any observer frame.
We label the momenta of the particles as follows:
the incoming fermion has momentum $q$,
the emitted gauge boson has momentum $p$,
and the outgoing fermion has momentum $q' = q-p$.
We assume that the fermions obey a conventional Lorentz-symmetric dispersion relation and that $m_1 < M$ as well as $m_2 < M$.
From simple kinematic considerations it follows that
the decay can only take place if $p^2 < (m_1-m_2)^2 < M^2$.
In the only polarization mode that allows this inequality to be satisfied, the $W$ bosons obey the dispersion relation \cite{covquant}
\begin{equation}
\Lambda_+(p) \equiv p^2 - M^2 + 2\sqrt{(p\cdot k)^2 - p^2 k^2} = 0\ ,
\label{dispersionrelation}
\end{equation}
where we dropped the subscript on $k_{2}^\mu$ for conciseness. Combining this with the aforementioned inequality $p^2 < (m_1-m_2)^2$, it follows that
\begin{equation}
4(p\cdot k)^2 > (M^2-(m_1 - m_2)^2)^2 + 4 k^2 (m_1 - m_2)^2
\label{decaycondition}
\end{equation}
is a necessary condition for the decay to be possible. For $m_1 = m_2$ this becomes the condition for the $W$-boson momentum to be spacelike, i.e. $p^2<0$ if and only if
\begin{equation}
4(p\cdot k)^2 > M^4\ .
\label{spacelikecondition}
\end{equation}
It is interesting to note that once such a gauge boson with spacelike momentum exists, it cannot decay to particles with exclusively timelike momenta.
This follows directly from energy-momentum conservation and the reversed triangle inequality for timelike vectors in Minkowski space. 
It might be interesting to investigate this in the context of cosmology and
the dark-matter content of the universe, however,
this lies outside the scope of the present considerations. 

\begin{figure}
\centering
\includegraphics[scale = 0.37]{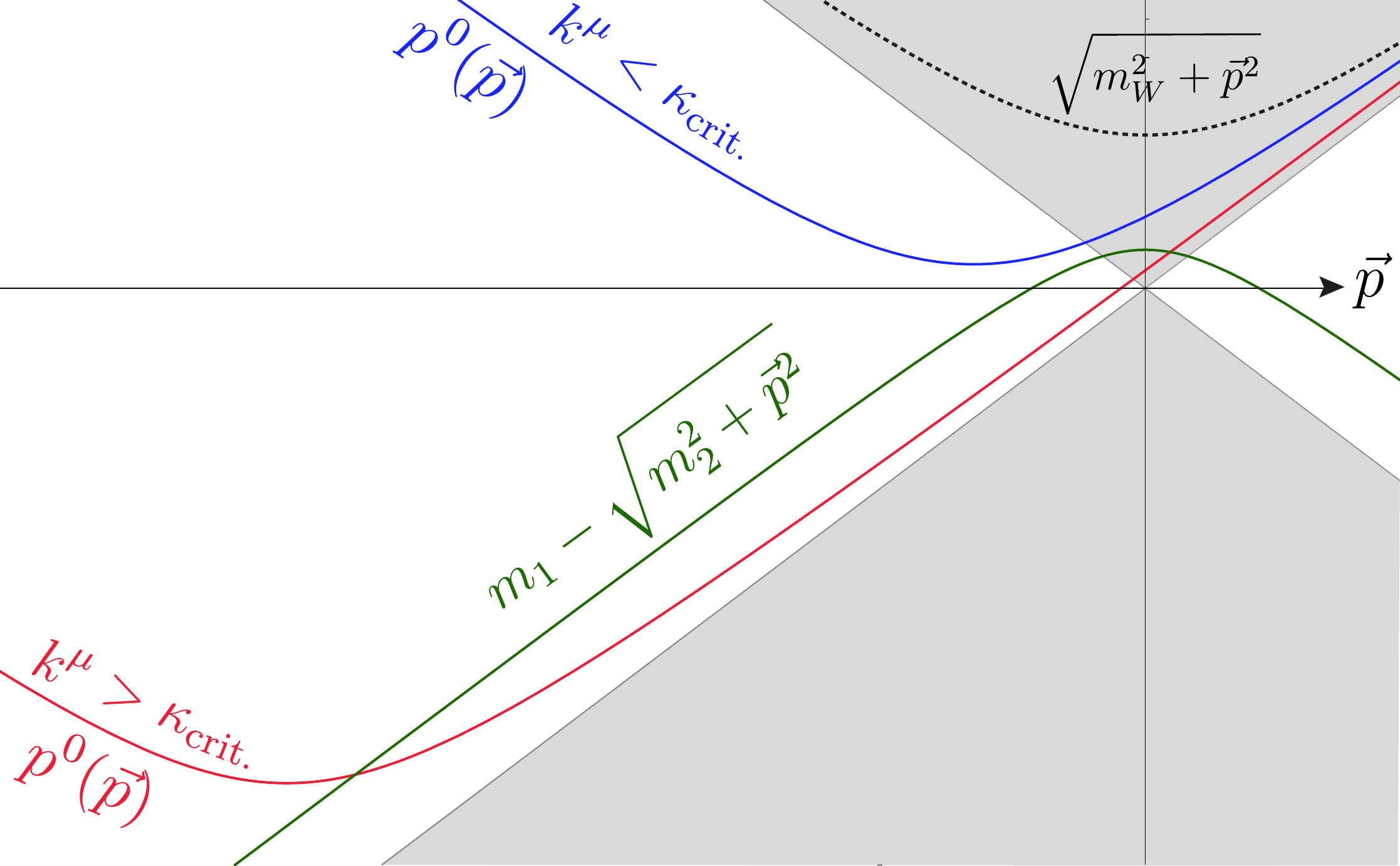}
\caption{The green line is the energy difference $q^0 - q^{\prime 0}$ in the rest frame of the incoming particle. The red (blue) line represents a possible solution of the $W$-boson dispersion relation for $k^\mu$ smaller (larger) than some critical value $\kappa_{\rm crit.}$. The gray area represents the lightcone for which $p^2 > 0$.} 
\label{fig:dispplot}
\end{figure}

That there exist, in fact, sets of $(q,p,q')$ in which dispersion relation
\eqref{dispersionrelation} for the $W$ momentum $p$ is satisfied,
can be seen fairly simply in the rest frame of the incoming fermion,
see Fig.~\ref{fig:dispplot}. In this frame the components of $k^\mu$ must be larger than some critical value $\kappa_{\rm crit.} \simeq \frac{M^2-(m_1-m_2)^2}{2m_1}$, i.e. the rest frame should be highly boosted with respect to so-called concordant frames, where $k^\mu$ is phenomenologically constrained to be small. This implies very large incoming momenta in concordant frames, cf. Eq.~\eqref{qthreshold}. In Fig.~\ref{fig:dispplot}, we used $m_1 > m_2$, which allows for a small range of timelike $p^\mu$ that satisfy the energy-momentum balance, as well as Eq.~\eqref{dispersionrelation}.   

We now turn to the differential decay rate, which is given by
\begin{align}
d\Gamma &= \frac{1}{2q^0}\frac{d^3p}{(2\pi)^3}\frac{1}{\Lambda'_+(p)}\frac{d^3q'}{(2\pi)^3}\frac{1}{2q^{\prime 0}} \left(\frac{1}{2}\sum_{\rm spins}|\mathcal{M}|^2\right)(2\pi)^4\delta^4(q - p -q')\ .
\label{relationmatrixdecayrate}
\end{align}
Here, the squared matrix element $|\mathcal{M}|^2$ is summed (averaged) over the final (initial) fermion spin.
The unconventional, but positive definite \cite{covquant}, factor
\begin{equation}
\Lambda'_+(p) = \frac{\partial \Lambda_+(p)}{\partial p^0}
\label{Lambda-prime}
\end{equation}
in the denominator defines a normalization in which the phase space and the matrix element are separately observer Lorentz invariant \cite{covquant},
i.e. invariant under simultaneous Lorentz transformations of the momenta and the LV four-vector.
Explicit observer Lorentz covariance of the formalism is important
because it will allow us to transform to convenient observer frames later on.
Note that the phase space normalization using (\ref{Lambda-prime}) even
allows transformations to observer frames in which the energy of the emitted
boson goes to zero,
typically leading to divergent factors in conventional formalisms for LV.

The matrix element that follows from the appropriate tree-level Feynman diagram is given by
\begin{equation}
i\mathcal{M} = \frac{ig}{2\sqrt{2}} \bar{u}(q')\gamma^\mu(1- \gamma^5)u(q)e^{(+)*}_\mu(p)\ ,
\label{current}
\end{equation}
where $u(q)$ and $u(q')$ are conventional Dirac spinors.
An analogous expression can be written down for the matrix element
for antiparticles.  
The four-vector $e^{(+)}_\mu(p)$ is the gauge-boson polarization vector
that corresponds to the dispersion relation in Eq.~\eqref{dispersionrelation}. 
The explicit expression for the latter can be found in Ref.~\cite{covquant}. 
The constant $g\simeq 0.65$ is the $SU(2)$ coupling constant.

Using the fact that \cite{covquant}
\begin{align}
&e^{(+)}_\mu(p)e^{(+)*}_\nu(p) = -\frac{1}{2}\eta_{\mu\nu}- \frac{p_\mu p_\nu k^2 + k_\mu k_\nu p^2 - (p_\mu k_\nu + p_\nu k_\mu)(p\cdot k)}{2((p\cdot k)^2 - p^2 k^2)}+ \frac{i\epsilon_{\mu\nu\alpha\beta}k^\alpha p^\beta}{2\sqrt{(p\cdot k)^2 - p^2 k^2}}\ ,
\end{align}
we find that the spin-summed squared matrix element is given by
\begin{equation}
\sum_{\rm spins}|\mathcal{M}|^2 = -\frac{g^2}{4}p^2\left[\left(1 \mp X\right)^2 - \frac{(m_1^2 - m_2^2)^2}{p^4} \right]\ ,
\end{equation}
where the upper (lower) sign holds for a decaying particle (antiparticle) and
\begin{equation}
X = \frac{p^2(p\cdot k - 2 q\cdot k) + (m_1^2 - m_2^2)(p\cdot k)}{p^2\sqrt{(p\cdot k)^2 - p^2 k^2}}\ .
\end{equation}
The decay rate becomes
\begin{align}
d\Gamma &= -\frac{g^2}{64 \pi^2 q^0}\int\frac{d^3p}{\Lambda'_+(p)}
\theta(q^0-p^0)\delta((q-p)^2-m^2) p^2\left[\left(1 \mp X\right)^2 - \frac{(m_1^2 - m_2^2)^2}{p^4} \right]\ .
\label{rateformula}
\end{align}
Here, the requirement that $p^2 < (m_1 - m_2)^2$ is reflected by the fact that it is a necessary condition for the product of the delta function and the step function to be non-vanishing. Moreover, the step function is automatically satisfied for momenta, $p^2 < (m_1 - m_2)^2$, for which the delta function has support.

Because of the observer-Lorentz-covariant normalization, it is easy to see that $d\Gamma$ transforms as $1/q^0$ under an observer Lorentz transformation. To perform the integrations over $\vec{p}$, given in Eq.~\eqref{rateformula}, we can thus specialize to an observer frame that simplifies the calculation. For the cases that $k^\mu$ is timelike or spacelike, we go the frame where $k^\mu$ is purely timelike, i.e. $k = (k^0,\vec{0})$, or purely spacelike, i.e. $k = (0,\vec{k})$, respectively. If $k^\mu$ is lightlike, we do not need to specialize to a particular frame. We assume that the components of $k^\mu$ are small, compared to $M$, in the selected frames. This is equivalent to assuming the existence of concordant frames, since in other frames than those with purely spacelike/timelike $k^\mu$, $k^\mu$ will have larger values for (some of) its components, while for lightlike $k^\mu$ we can always find a frame where the components are small.

In the mentioned frames Eq.~\eqref{decaycondition} translates to
\begin{equation}
|\vec{p}| > \frac{\tilde{M}^2}{2|\kappa|}\ ,
\end{equation}
where $\tilde{M}^2 = M^2 + \mathcal{O}\left(\tfrac{\kappa^2}{M^2},\tfrac{m^2_{1,2}}{M^2}\right)$ and $\kappa = k^0$ for purely timelike $k^\mu$,
$\kappa = |\vec{k}|(1- {\rm sgn}(k^0)\cos\theta_{pk})$ for lightlike $k^\mu$,
and $\kappa = |\vec{k}|\cos\theta_{pk}$ for purely spacelike $k^\mu$
($\theta_{pk}$ denotes the angle between the vectors indicated by the subscript).
Furthermore, the delta function in Eq.~\eqref{rateformula} demands that
\begin{equation}
\cos\theta_{pq} = 1 + \mathcal{O}(\kappa^2/M^2)\ .
\label{anglepq}
\end{equation}
All gauge bosons are thus emitted in a very narrow forward beam around the direction of the incoming fermion. It also follows that $\cos\theta_{pk} = \cos\theta_{qk}$, up to terms quadratic in LVCs. 

The fact that $\cos^2\theta_{pq}$ has to be smaller than unity determines the integration limits for $|\vec{p}|$.
They are found by demanding that the higher-order terms in Eq.~\eqref{anglepq} are negative.
Although straightforward to obtain,
the explicit expressions for the integration limits are not very illuminating.
They can be approximated by
\begin{equation}
|\vec{p}|_{\rm min} \approx  \frac{M^2}{2|\kappa|} \qquad {\rm and}\qquad |\vec{p}|_{\rm max} \approx  |\vec{q}|\ ,
\label{integrationlimits}
\end{equation}
where we omitted terms of order $\kappa^2/M^2$ and terms quadratic in the fermion to gauge-boson mass ratio. Within the same approximation, the threshold value for $|\vec{q}|$ is given by
\begin{equation}
|\vec{q}|_{\rm th} = \frac{M(M+2m_2)}{2|\kappa|}\ ,
\label{qthreshold}
\end{equation}
meaning that fermions with an absolute momentum larger than $|\vec{q}|_{\rm th}$ will start decaying while emitting a $W$ boson.

Using Eqs.~\eqref{anglepq} and \eqref{integrationlimits}, we can perform the integration over $\vec{p}$. We find that the decay rate is given by
\begin{equation}
\Gamma =\frac{g^2 |\kappa|}{64 \pi}  G(a)\theta(a -1) .
\label{decayrate3a}
\end{equation}
where
\begin{align}
G(a) &= \alpha(a)\bigg[\frac{1}{a}(-7 + 3y)\left(1 - \frac{m_2}{M}\right) - \frac{1}{a^2}(1 - y)\left(1 - \frac{3m_2}{M}\right)\bigg] \nonumber \\
&\quad
{} - 4\left(1 + \frac{1}{a}(1 - y)\right)\left(1 - \frac{2m_2}{M}\right) \log\left(\frac{1 + a - m_2(1-a)/M - \alpha(a)}{1 + a - m_2(1-a)/M + \alpha(a)}\right) \nonumber \\
&\qquad\quad
{}+ \mathcal{O}\left(\frac{m_{1,2}^2}{M^2},\frac{\kappa^2}{M^2}\right)
\label{Gexplicit}
\end{align}
with %$a = E/E_{\rm th}$ and 
$\alpha(a) = \sqrt{(a-1)^2+2m_2(a^2-1)/M}$ and
$y = \pm\mathrm{sgn}(\kappa)$, where the upper (lower) sign applies to the particle (antiparticle).

The variable $a$ is defined as the ratio of $|\vec{q}|$ to its threshold value, i.e. $a = |\vec{q}|/|\vec{q}|_{\rm th}$. In a general observer frame and up to terms of order $\kappa^2 / M^2$ and $m_{1,2}^2 / M^2$, we can write this as
\begin{equation}
a = \frac{2|q\cdot k|}{M(M+2m_2)}\ .
\end{equation}
The step function in Eq.~\eqref{decayrate3} imposes the threshold condition for the initial fermion.

A fermion that interacts with a CPT-violating $W$ boson will start emitting $W$ bosons if the fermion has an energy above threshold. The $W$ boson will carry away at least an energy that corresponds to Eq.~\eqref{qthreshold}. From Eq.~\eqref{decayrate3a} we determine that the typical decay time is in the order of $10^{-15}\ {\rm s}$ if $\mathcal{O}(\kappa) = 10^{-7}\ {\rm GeV}$, corresponding to the bound we will find later on. This means that it will take about $a \times 10^{-15}\ {\rm s}$ for all fermions in a decay cascade to fall below threshold for such values of $\kappa$.

Strictly speaking, these results are only valid for elementary fermions and not for composite particles.
In the important case of the proton, the emission of a $W$ boson will provoke a break-up, since the typical momentum transfer lies in the range of the $W$-boson mass, which is well within the energy range of for example deep inelastic scattering. In this case, the proton-decay rate can be written as
\begin{equation}
\Gamma = \frac{1}{2q^0}\int \frac{d^3\vec{p}}{(2\pi)^3}\frac{4\pi}{\Lambda_+'(p)}e^{(+)}_\mu (p)e^{(+)*}_\nu (p)W^{\mu\nu}\ ,
\label{decayrate1}
\end{equation}
where $W^{\mu\nu}$ is the hadronic part, given by
\begin{equation}
W^{\mu\nu} = \frac{1}{8\pi}\sum_\sigma\langle {\rm p}(q,\sigma)|J^\nu(-p)\SumInt_X |X\rangle\langle X|J^\mu(p)|{\rm p}(q,\sigma)\rangle
\end{equation}
while $|{\rm p}(q,\sigma)\rangle$ is a proton state with momentum $q$ and spin $\sigma$, $J^\mu(p)$ is the hadronic current, and $\SumInt_X$ represents a sum over all hadronic final states $X$ along with the corresponding integrations over phase space. $W^{\mu\nu}$ can be evaluated in the parton 
model. This essentially involves calculating the decay rate of an elementary quark that caries a fraction $x$ of the longitudonal proton momentum. We can thus use many of the results obtained for the elementary fermion rate. For a pedagogical introduction to parton-model calculations, we refer to Ref.~\cite{PeskinSchroeder}.

The final result for the decay rate is
\begin{equation}
\Gamma =\frac{g^2|\kappa|}{64 \pi}\sum_q \int_0^1 dx\,(f_q(x) + \bar{f}_q(x))
\tilde{G}_q(ax)\theta(ax -1)\ .
\label{decayrate3}
\end{equation}
Here the functions $f_q(x)$ and $\bar{f}_q(x)$ are the parton distributions functions (PDFs) for the quarks and antiquarks of flavor $q$, respectively. They represent the chance of finding a quark with momentum fraction $x$ inside the proton. We assumed the PDFs to be independent of $p^2$, which is a good approximation to leading order in the strong coupling constant. The function $\tilde{G}_q(ax)$ in Eq.~\eqref{decayrate3} is the function in Eq.~\eqref{Gexplicit} with the substitutions $m_2 \rightarrow x m_2$ and $y \rightarrow \tilde{y}_q = {\rm sgn}(\kappa)\frac{f_q(x) - \bar{f}_q(x)}{f_q(x) + \bar{f}_q(x)}$. As expected, Eq.~\eqref{decayrate3} is basically the sum over elementary-quark-decay rates, weighted by the relevant PDF.

The integral over $x$ in Eq.~\eqref{decayrate3} can be carried out numerically
using fits for the PDFs \cite{pdf}. The presence of the PDFs is particularly important for energies close to threshold, i.e. when $a\approx 1$. Here, $x$ also has to be close to one for the decaying quark to be above threshold, i.e. $ax > 1$. At such large values of $x$, the proton PDFs for valence quarks decay to zero approximately as a constant times $(1-x)^{c_q}$ with $c_u \approx 4$ and $c_d \approx 5$, typically \cite{pdf}.
For this reason, the integral over $x$ in Eq.~\eqref{decayrate3} yields
decay rates just above threshold that are considerably smaller than if the proton would have been an elementary particle.

To obtain a bound on $\kappa$, we observe that a cosmic-ray proton that has an energy below threshold has zero probability
to disintegrate by $W$-boson emission, and can thus reach Earth unimpeded.
Above threshold, the proton can disintegrate, and thus it cannot reach Earth
if its mean free path is much smaller than the distance $D$ from its source
to Earth.

Since many ultra-high-energy cosmic-ray (UHECR) particles with energies
above $57\,\mathrm{EeV}\equiv |\vec{q}|_{obs}$ have been observed,
coming more or less from all directions \cite{uhecrdetected},
we can take it as a first estimate for the lower bound for $E_{\mathrm{th}}$.
It follows that
\begin{equation}
|\kappa| < \frac{M(M + 2m_2)}{|\vec{q}|_{obs}}
\approx 1.1 \times 10^{-7}\,\mathrm{GeV}\equiv|\kappa|_0\ .
\label{kappabound}
\end{equation}
This bound can only be relaxed if the mean free path of protons above
threshold is not much smaller than $D$.
From Eq.~\eqref{decayrate3} we see
that the mean lifetime of protons (in Earth's frame) $t_p$
is still proportional to $|\kappa|^{-1}$, but is enhanced, mainly by the minute values of the PDFs at large $x$.
A conservative estimation that comes from comparing to the elementary-fermion decay time gives a mean free path of
\begin{equation}
L \simeq c t_p \sim (\hbar c/|\kappa|_0)\times 10^{15} \approx 10^3\ {\rm km}\ .
\label{L}
\end{equation}
Clearly, in such a scenario, protons with an energy above threshold
will not be able to reach Earth from any viable UHECR source.
We thus obtain a bound on all four components of the LVC:
\begin{equation}
|k_{2}^\mu| < 1.1 \times 10^{-7}\ {\rm GeV}\ .
\label{limitresult}
\end{equation}

We note that we have to assume that at least some of the detected UHECRs are protons and that these have a sufficient spread in arrival direction. Although the mass content of the UHECRs, in particular at high energies, is still largely unexplored \cite{uhecrmass}, it seems very unlikely that such a significant low-mass component is completely absent. Moreover, even if this is the case, one can calculate a decay rate equivalent to Eq.~\eqref{decayrate3}, by using nuclear PDFs \cite{npdf}. Since $L \ll D$ by many orders of magnitude, it is highly improbable that any of this will change our result in Eq.~\eqref{limitresult}.

Using the limit in Eq.~\eqref{limitresult}
and the fact that the photon parameter
$k_{AF}^\mu = 2\cos^2\theta_w k_1^\mu + \sin^2\theta_w k_2^\mu$
is bounded to be virtually zero \cite{datatables},
we find bounds on the fundamental parameters $k_1^\mu$ and $k_2^\mu$,
given by $|k_1^\mu| < 1.7 \times 10^{-8}\,\mathrm{GeV}$ and
$|k_2^\mu| < 1.1 \times 10^{-7}\,\mathrm{GeV}$.
This thus limits the entire CPT-odd $SU(2)\times U(1)$ gauge sector
of the mSME to be smaller than about $10^{-7}\,\mathrm{GeV}$.

\acknowledgments
This work is supported in part by the Funda\c c\~ao para a Ci\^encia e a Tecnologia of Portugal (FCT) through projects UID/FIS/00099/2013 and SFRH/BPD/101403/2014 and program POPH/FSE.


\begin{thebibliography}{99}

\bibitem{qgmodels}
  V.~A.~Kosteleck\'y and S.~Samuel,
  %``Spontaneous Breaking of Lorentz Symmetry in String Theory,''
  Phys.\ Rev.\ D {\bf 39}, 683 (1989);
  %%CITATION = doi:10.1103/PhysRevD.39.683;%%
  V.~A.~Kostelecky and R.~Potting,
  %``CPT and strings,''
  Nucl.\ Phys.\ B {\bf 359}, 545 (1991).
  %%CITATION = doi:10.1016/0550-3213(91)90071-5;%%
  J.~R.~Ellis, N.~E.~Mavromatos, and D.~V.~Nanopoulos,
  %``Search for quantum gravity,''
  Gen.\ Rel.\ Grav.\  {\bf 31}, 1257 (1999);
  %%CITATION = doi:10.1023/A:1026720723556;%%
  R.~Gambini and J.~Pullin,
  %``Nonstandard optics from quantum space-time,''
  Phys.\ Rev.\ D {\bf 59}, 124021 (1999);
  %%CITATION = doi:10.1103/PhysRevD.59.124021;%%
  C.~P.~Burgess, J.~M.~Cline, E.~Filotas, J.~Matias, and G.~D.~Moore,
  %``Loop generated bounds on changes to the graviton dispersion relation,''
  JHEP {\bf 0203}, 043 (2002).
  %%CITATION = doi:10.1088/1126-6708/2002/03/043;%%

\bibitem{sme}
  D.\ Colladay and V.~A.\ Kosteleck\'y,
  Phys.\ Rev.\ D {\bf 55}, 6760 (1997);
 %%CITATION = doi:10.1103/PhysRevD.55.6760;%%
  Phys.\ Rev.\ D {\bf 58}, 116002 (1998).
 %%CITATION = doi:10.1103/PhysRevD.58.116002;%%

\bibitem{gravity}
  V.~A.~Kosteleck\'y,
  %``Gravity, Lorentz violation, and the standard model,''
  Phys.\ Rev.\ D {\bf 69}, 105009 (2004).
  %%CITATION = doi:10.1103/PhysRevD.69.105009;%%

\bibitem{Gre02}
  O.~W.~Greenberg,
  %``CPT violation implies violation of Lorentz invariance,''
  Phys.\ Rev.\ Lett.\  {\bf 89}, 231602 (2002);
  %%CITATION = doi:10.1103/PhysRevLett.89.231602;%%

\bibitem{datatables}
  V.~A.~Kostelecky and N.~Russell,
  %``Data Tables for Lorentz and CPT Violation,''
  Rev.\ Mod.\ Phys.\  {\bf 83}, 11 (2011)
  [2016 edition: arXiv:0801.0287v9 [hep-ph]].
  %%CITATION = doi:10.1103/RevModPhys.83.11;%%

\bibitem{vacuumchernk}
  R. Lehnert and R. Potting, Phys. Rev. Lett. {\bf 93}, 110402 (2004); Phys. Rev. D {\bf 70}, 125010 (2004); erratum {\it ibid}. {\bf 70}, 129906
  (2004);  C. Kaufhold and F. R. Klinkhamer, Nucl. Phys. B {\bf 734}, 1 (2006); B. Altschul, Phys. Rev. Lett. {\bf 98}, 041603 (2007). D. Colladay, P. 
  McDonald, and R. Potting, Phys. Rev. D {\bf 93}, 125007 (2016).

\bibitem{Carroll} 
  S.~M.~Carroll, G.~B.~Field, and R.~Jackiw,
  %``Limits on a Lorentz and Parity Violating Modification of Electrodynamics,''
  Phys.\ Rev.\ D {\bf 41}, 1231 (1990).
  %%CITATION = doi:10.1103/PhysRevD.41.1231;%%

\bibitem{covquant}
  D.~Colladay, P.~McDonald, J.~P.~Noordmans, and R.~Potting,
  %``Covariant Quantization of CPT-violating Photons,''
  Phys.\ Rev.\ D {\bf 95}, 025025 (2017).
  %%CITATION = doi:10.1103/PhysRevD.95.025025;%%

\bibitem{kphicontr}
  Naively, one expects a CPTV contribution to the $SU(2)\times U(1)$ gauge sector from the CPT-odd Higgs-sector coefficient $k_\phi^\mu$ \cite{sme}. However, to first order in LV this contribution vanishes due to a cancellation between the tree-level Lagrangian and the vacuum expectation value for $Z$, that is induced by $k_\phi^\mu$.

\bibitem{mixingnote}
  At low energy, the photon $Z$-boson mixing term may be removed to first order in LV by a field redefinition. However, at the energies that are presently relevant, i.e. energies such that $p\cdot k \sim M^2$, such a procedure fails. This is the reason we consider the $W$-boson instead of the $Z$-boson in this work.
  
\bibitem{pdg} 
  C.~Patrignani {\it et al.} [Particle Data Group],
  %``Review of Particle Physics,''
  Chin.\ Phys.\ C {\bf 40}, no. 10, 100001 (2016).
  %%CITATION = doi:10.1088/1674-1137/40/10/100001;%%

\bibitem{PeskinSchroeder} 
  M.~E.~Peskin and D.~V.~Schroeder,
  {\it An Introduction to quantum field theory\/}, (Addison-Wesley, 1995).
  
\bibitem{pdf}
  D.~Stump, J.~Huston, J.~Pumplin, W.~K.~Tung, H.~L.~Lai, S.~Kuhlmann, and J.~F.~Owens,
  %``Inclusive jet production, parton distributions, and the search for new physics,''
  JHEP {\bf 0310}, 046 (2003);
  %%CITATION = doi:10.1088/1126-6708/2003/10/046;%%
  S.~Alekhin, K.~Melnikov, and F.~Petriello,
  %``Fixed target Drell-Yan data and NNLO QCD fits of parton distribution functions,''
  Phys.\ Rev.\ D {\bf 74}, 054033 (2006);
  %%CITATION = doi:10.1103/PhysRevD.74.054033;%%
  J.~F.~Owens, J.~Huston, C.~E.~Keppel, S.~Kuhlmann, J.~G.~Morfin, F.~Olness, J.~Pumplin, and D.~Stump,
  %``The Impact of new neutrino DIS and Drell-Yan data on large-x parton distributions,''
  Phys.\ Rev.\ D {\bf 75}, 054030 (2007);
  %%CITATION = doi:10.1103/PhysRevD.75.054030;%%
  
\bibitem{uhecrdetected} 
  A.~Aab {\it et al.} [Pierre Auger Collaboration],
  %``Searches for Anisotropies in the Arrival Directions of the Highest Energy Cosmic Rays Detected by the Pierre Auger Observatory,''
  Astrophys.\ J.\  {\bf 804}, no. 1, 15 (2015);
  %%CITATION = doi:10.1088/0004-637X/804/1/15;%%
  T.~Abu-Zayyad {\it et al.} [Telescope Array Collaboration],
  %``Correlations of the Arrival Directions of Ultra-high Energy Cosmic Rays with Extragalactic Objects as Observed by the Telescope Array Experiment,''
  Astrophys.\ J.\  {\bf 777}, 88 (2013).
  %%CITATION = doi:10.1088/0004-637X/777/2/88;%%

\bibitem{uhecrmass} 
  A.~Aab {\it et al.} [Pierre Auger Collaboration],
  %``Depth of maximum of air-shower profiles at the Pierre Auger Observatory. II. Composition implications,''
  Phys.\ Rev.\ D {\bf 90}, no. 12, 122006 (2014);
  %%CITATION = doi:10.1103/PhysRevD.90.122006;%%
  R.~U.~Abbasi {\it et al.},
  %``Study of Ultra-High Energy Cosmic Ray composition using Telescope Array’s Middle Drum detector and surface array in hybrid mode,''
  Astropart.\ Phys.\  {\bf 64}, 49 (2015)
  %%CITATION = doi:10.1016/j.astropartphys.2014.11.004;%%


\bibitem{npdf} 
  K.~Kovarik {\it et al.},
  %``nCTEQ15 - Global analysis of nuclear parton distributions with uncertainties in the CTEQ framework,''
  Phys.\ Rev.\ D {\bf 93}, no. 8, 085037 (2016)
  [arXiv:1509.00792 [hep-ph]].
  %%CITATION = doi:10.1103/PhysRevD.93.085037;%%
  
\end{thebibliography}
\end{document}